# Pauli-limit violation in lanthanide infinite-layer nickelate superconductors

L. E. Chow,[1] K. Y. Yip,[2] M. Pierre,[3] S. W. Zeng,[1] Z. T. Zhang,[1] T. Heil,[4] J. Deuschle,[4] P. Nandi,[1] S. K. Sudheesh,[1] Z. S. Lim,[1] Z. Y. Luo,[1] M. Nardone,[3] A. Zitouni,[3] P. A. van Aken,[4] M. Goiran,[3] S. K. Goh,[2] W. Escoffier,[3] A. Ariando[1,✉]

[1]Department of Physics, Faculty of Science, National University of Singapore, Singapore 117551, Singapore

[2]Department of Physics, The Chinese University of Hong Kong, Shatin N.T., Hong Kong SAR, China

[3]LNCMI, Université de Toulouse, CNRS, INSA, UPS, EMFL, 31400 Toulouse, France

[4]Max Planck Institute for Solid State Research, 70569 Stuttgart, Germany

✉To whom correspondence should be addressed: ariando@nus.edu.sg

# Abstract

**Superconductivity can be destroyed by a magnetic field with an upper bound known as the Pauli-limit in spin-singlet superconductors.[1,2] Almost all the discovered superconductors are spin-singlet, with the highest transition temperature $T_c$ at ambient pressure achieved in the cuprate family.[3,4] The closest cuprate analogue is the recently discovered infinite-layer nickelate, which hosts substantial structural and electronic similarity to the cuprate.[5–10] A previous magnetotransport study on $Nd_{0.775}Sr_{0.225}NiO_2$ has observed an isotropic Pauli-limited upper critical field.[11] Here, we report a large violation (> 2 times) of Pauli-limit in every crystallographic direction in $La_{1-x}(Ca/Sr)_xNiO_2$ regardless of the doping $x$. Such a large violation of the Pauli-limit in all directions in $La_{1-x}(Ca/Sr)_xNiO_2$ is unexpected and unlikely accounted by a Fulde-Ferrell-Larkin-Ovchinnikov (FFLO)-state,[12] strong spin-orbit-coupling,[13,14] strong-coupling or a large pseudogap. On the other hand, in agreement with the previous report, we observe a Pauli-limiting critical field in $Nd_{1-x}Sr_xNiO_2$ and the superconducting anisotropy decreases as doping increases, suggesting a spin-singlet pairing. Therefore, superconductivity in $La_{1-x}(Ca/Sr)_xNiO_2$ could be driven by a non-spin-singlet Cooper pairing mechanism with an attractive 'high-$T_c$' at 10 K, an order of magnitude higher than the known spin-triplet superconductors, favourably extending the application of spin-triplet superconductivity in topological matter, non-dissipative spintronics, and quantum computing.[15–17]**

# Main text

The discovery of high critical transition temperature (high-$T_c$) superconductivity in copper oxide-based compounds (cuprates) has reshaped the landscape of strongly-correlated systems with broad technological prospects. As one of the biggest unsolved mysteries in condensed matters in the past three decades,[3,18] the mechanism of high-$T_c$ superconductivity in the cuprate has been heavily investigated with intense effort in the search for 'cuprate-analogue' compounds that can mimic the cuprate's structural and electronic design, to decipher the secret ingredient of high-$T_c$ superconductors.[5–9] The recently discovered infinite-layer nickelate, which shares the closest electronic structure to the cuprate, has been long promised to be the ideal sister, which shall mirror the cuprate's key properties.[7–9,19–24] Some of these properties have been experimentally demonstrated, for example, charge order and magnons in the undoped and underdoped regime,[25–28] strange metal behaviour near optimal doping,[29,30] and Fermi-liquid behaviour in the overdoped regime.[30] On the other hand, several distinctive features between infinite-layer nickelate and cuprate have been illustrated: the Mott-Hubbard physics instead of charge-transfer insulator parent compound,[6,9,23] not a single $d$-wave gap,[21,31,32] and isotropic upper critical fields in overdoped $Nd_{0.775}Sr_{0.225}NiO_2$ in contrast to the anisotropic nature of quasi-two-dimensional layered structure.[11]

Despite considerable advances in understanding superconductivity in nickelates, the most important question remains an assumption – the spin and parity nature of the order parameter. Superconductivity arises from the pairing of electrons into Cooper pairs, which can be either spin-singlet or spin-triplet. The parity decides the possible spatial symmetry, such as $s$-wave or $d$-wave for even parity, which is typical for spin-singlet pairing, and odd-parity $p$-wave, that is common for spin-triplet superconductors, which recently meet exciting interests in quantum computing and non-dissipative spintronics.[15–17] As a cuprate analogue, one assumes nickelate to follow spin-singlet pairing with even parity. However, the correctness of this assumption is critical, since it defines the possible spatial symmetry and, fundamentally, the superconducting mechanism. A recent penetration depth study has proposed distinctive order parameters between Nd- and La- infinite-layer nickelates,[31] however, possible options of order parameter can only be constructed if employing the spin-singlet assumption.[31,32] Insights into the spin configuration of the pairing symmetry – singlet or triplet – can be obtained from the upper critical fields $H_{c2}$ measurement of a superconductor. In a spin-singlet superconductor, a sufficiently large magnetic field can destroy the Cooper pairs through two mechanisms: orbital effect (formation of vortices in type-II superconductors) and Pauli paramagnetic de-pairing effect (Zeeman effect). The later imposes an upper bound to the $H_{c2}$ for a spin-singlet superconductor known as the Pauli (Clogston-Chandrasekhar)-limit[1,2] $H_p$, which characterises the binding energy of Cooper pairs. For a weak-coupled BCS superconductor and $g$-factor of 2, the Pauli-limit $H_p = 1.84 \text{ T/K} \times T_c$. A spin-singlet superconductor, including an unconventional superconductor like a cuprate, follows $H_{c2} < H_p$, except in the presence of finite-momentum pairing (which leads to Fulde-Ferrell-Larkin-Ovchinnikov state),[12] strong spin-orbit-coupling (which locks the spins of Cooper pairs in a direction and is protected against orthogonal magnetic field),[13,14] large pseudogap or strong-coupling scenarios.

Here, we performed magnetotransport measurement on the infinite-layer nickelates Nd$_{1-x}$Sr$_x$NiO$_2$ and La$_{1-x}$(Ca/Sr)$_x$NiO$_2$ thin films using both pulsed-field up to 55 T and static field up to 14 T at temperature down to 0.03 K in a dilution fridge. For Nd-nickelate, agreeing with the earlier report on Nd$_{0.775}$Sr$_{0.225}$NiO$_2$,[11] we observed upper critical fields in both $c$-axis (out-of-plane) and $ab$-axis (in-plane) directions $H_{c2}^c, H_{c2}^{ab} < H_p$ at both $x = 0.15$ and $x = 0.2$ dopings, suggesting spin-singlet pairing. In addition, we found that the anisotropy in $H_{c2}$ decreases as hole doping increases, opposite to the picture of a more single-band, two-dimensional like electronic structure. The critical result of the present study is that superconductivity in the La- infinite-layer nickelate system survives in a very large magnetic field in both in-plane and out-of-plane directions, which $H_{c2} > H_p$ by 2 – 3 times, despite its lower $T_c$ as compared to the Nd-counterpart. Such a large violation of the Pauli-limit in all crystallographic directions, as discussed below, cannot be explained by typical mechanisms that allow superconductivity in a spin-singlet system to survive in magnetic fields beyond the Pauli-limit. Therefore, it is reasonable to propose a non-spin-singlet or mixed singlet-triplet pairing mechanism in the La- infinite-layer nickelate. With a 'high-$T_c$' of ~10 K that is one order of magnitude higher than the existing low $T_c \sim 1$ K in spin-triplet superconductors,[15] La-nickelate can be proven to be an attractive platform to navigate topological superconductivity and improve accessibility for non-dissipative spintronics and quantum computing research.[16,17] Our finding also suggests the richness of unconventional superconductivity in the infinite-layer nickelate family with large tunability to access various pairing landscapes through rare-earth dependency.

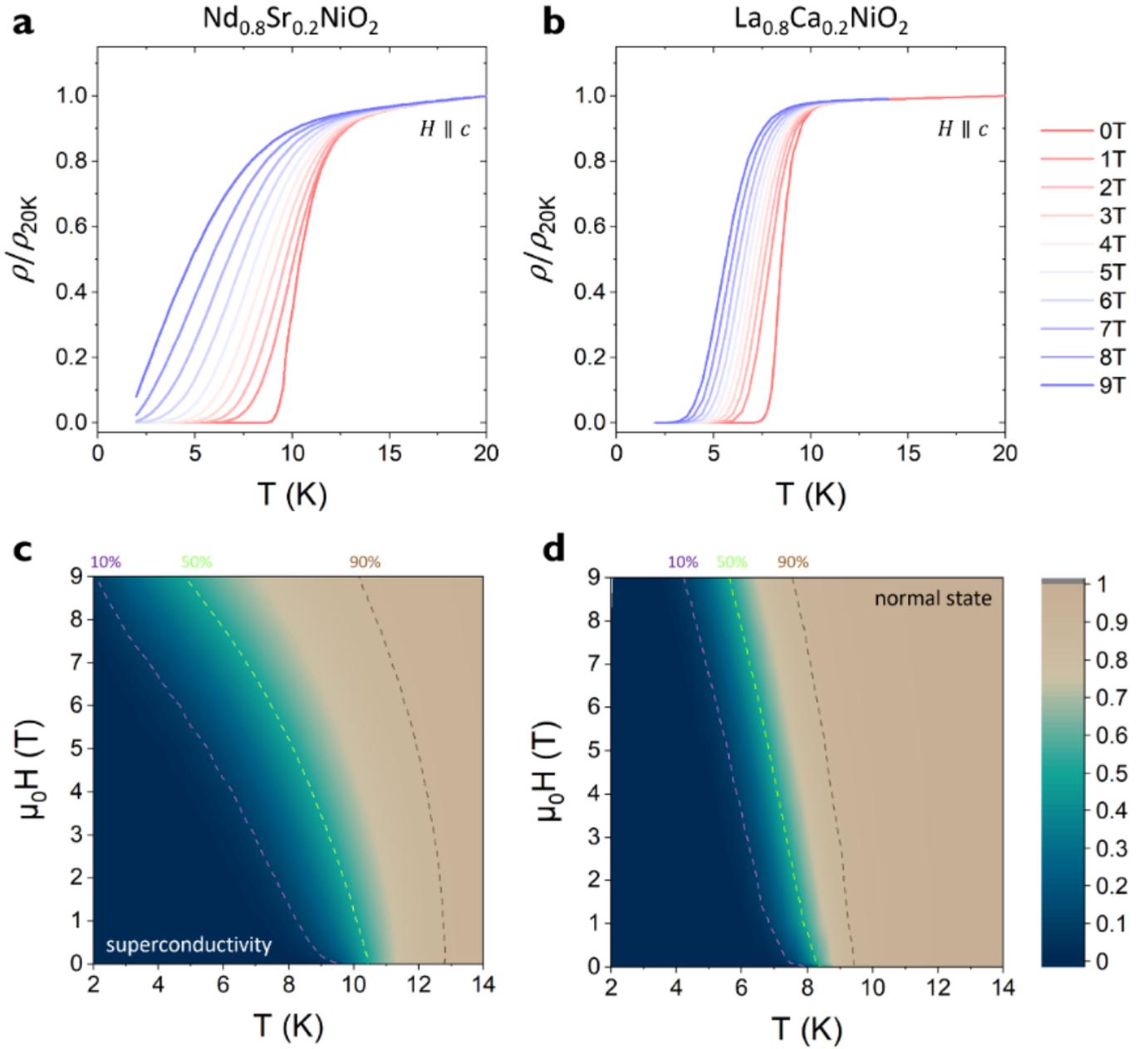

**Figure 1: Resistivity transitions of the superconducting Nd- and La- infinite-layer nickelate thin films under static magnetic field. (a-b)** $R-T$ curves at magnetic fields applied along the $c$-axis direction, for **(a)** $Nd_{0.8}Sr_{0.2}NiO_2$ and **(b)** $La_{0.8}Ca_{0.2}NiO_2$ thin films. **(c-d)** $H-T$ phase diagrams of the **(c)** Nd- and **(d)** La- nickelate thin films. Upper critical fields as a function of temperature are plotted for the 10%, 50% and 90% resistivity thresholds. Upon applied magnetic fields, obvious transition broadening is seen in the Nd-nickelate samples, the separation between 10% and 90% lines is wider as magnetic field increases. In contrast, the superconducting transition of the La- nickelate remains narrow from 1 T to 9 T, and robust against the destroy of superconductivity.

# Magnetotransport behaviour and anisotropy of infinite-layer nickelate

As previously reported,[31] high crystallinity infinite-layer nickelate samples are prepared without an *in-situ* SrTiO$_3$ capping layer. Cross-sectional scanning transmission electron microscopy high-angle dark-field (STEM-HAADF) images are shown in Extended Data **Fig. S2**. Resistivity transitions under applied magnetic fields along out-of-plane direction $H \parallel c$ of the superconducting Nd- and La- infinite-layer nickelate thin films are shown in **Fig. 1**. Apparent transition broadening is observed in Nd- nickelate samples, where the transition width $\Delta T_{90\% \rightarrow 10\%}$ increases drastically as the magnetic field increases. On the other hand, La- nickelate samples have a narrow transition, which is almost insensitive to the magnitude of applied magnetic fields. Such distinction in transition width between the two rare-earth systems suggests that the observed transition broadening in Nd-nickelate is not related to the static inhomogeneity or sample imperfection but is more likely a signature of the different vortex dynamics in the two systems. Given the observation of the robustness of the superconducting state against magnetic fields in La-nickelate despite the lower $T_c$ as compared to the Nd-counterpart, we extended the magnetotransport measurements to a magnetic field as high as 55 T and temperature down to 0.45 K on a La$_{0.8}$Ca$_{0.2}$NiO$_2$ thin film of $T_{c,0} \sim 7.3$ K. **Figures 2a,b** show the normalised resistivity as a function of magnetic field applied respectively along the $c$-axis and $ab$-axis directions at various temperatures. Overall, the transition widths $\Delta H_{0 \rightarrow 90\%}$ in both out-of-plane and in-plane directions are narrower as temperature decreases. A similar observation was made in the Nd$_{0.85}$Sr$_{0.15}$NiO$_2$ thin film (see Extended Data **Fig. S1**). Contrary to the earlier report on a Nd$_{0.775}$Sr$_{0.225}$NiO$_2$ thin film, where transitions are broadened as temperature decreases and an 'anomalous upturn' in $H_{c2}$ with large positive curvature appears at low temperature < 3 K or ~ 0.35 $T_c$,[11] we did not observe compelling evidence of a similar upturn of $H_{c2}$ in the Nd$_{1-x}$Sr$_x$NiO$_2$ thin film, even at a temperature down to 0.03 K (**Fig. 3**). Considering the thermodynamic instability of the system and hence extreme challenges in sample synthesis, we do not speculate on any explanation for the distinctive observation between ours and the earlier report.[11]

**Figures 2c,d** plot the $H - T$ phase diagram of the La$_{0.8}$Ca$_{0.2}$NiO$_2$ thin film. Considering the quasi-two-dimensional electronic structure and thin-film geometry, one can expect a large anisotropy in upper critical fields between $H \parallel c$ and $H \parallel ab$ directions. In the Nd-nickelate (see **Fig. 3** and **Fig. S1**), we do observe anisotropic $H_{c2}$ behaviour to a certain extent. For instance, at 4.2 K (**Figs. S1c-g**), the $H_{c2}^{ab}$ can be ~1.6 times larger than $H_{c2}^{c}$. As temperature decreases and when using a larger resistivity threshold, the anisotropy decreases, similar to the report on Nd$_{0.775}$Sr$_{0.225}$NiO$_2$.[11] In addition, we observed that the anisotropy decreases as Sr doping increases in the Nd-system (**Fig. 3**), suggesting a more isotropic magnetic field response in the large hole-doping regime. While the Ni $3d_{x^2-y^2}$ hole-band becomes more dominant and $c$-axis lattice constant increases as Sr doping increases, the observed doping dependency seems opposite to the expectation of a more two-dimensional like electronic structure at large hole doping. It was proposed that a more 'cuprate-like' nickelate can be engineered by truncating the $c$-axis dimensionality,[33] by, for example, increasing the separation between NiO$_2$ planes. In comparison, La-nickelate has a larger lattice constant due to the larger La$^{3+}$ ions, however, the anisotropy between $H_{c2}^{ab}$ and $H_{c2}^{c}$ is still relatively small, only ~1.3 times difference at $T \to 0$ K (**Figs. 2c,d**). For the temperature dependence of $H_{c2}$ in La- nickelate, $H_{c2}^{c}(T)$ shows a $T$-linear dependence (**Fig. 2c**) while $H_{c2}^{ab}(T) \propto \sqrt{1 - T/T_c}$ (**Fig. 2d**) down to the lowest temperatures. The $H_{c2}$ data at 10% and 50% resistivity thresholds both show a good agreement to the fitted dashed lines following the above relations. Interestingly, the $H_{c2}$ of La- nickelate shows no sign of saturation even at temperatures below 0.1 $T_c$, indicating absence of a Pauli-limiting ('flattening') behaviour.

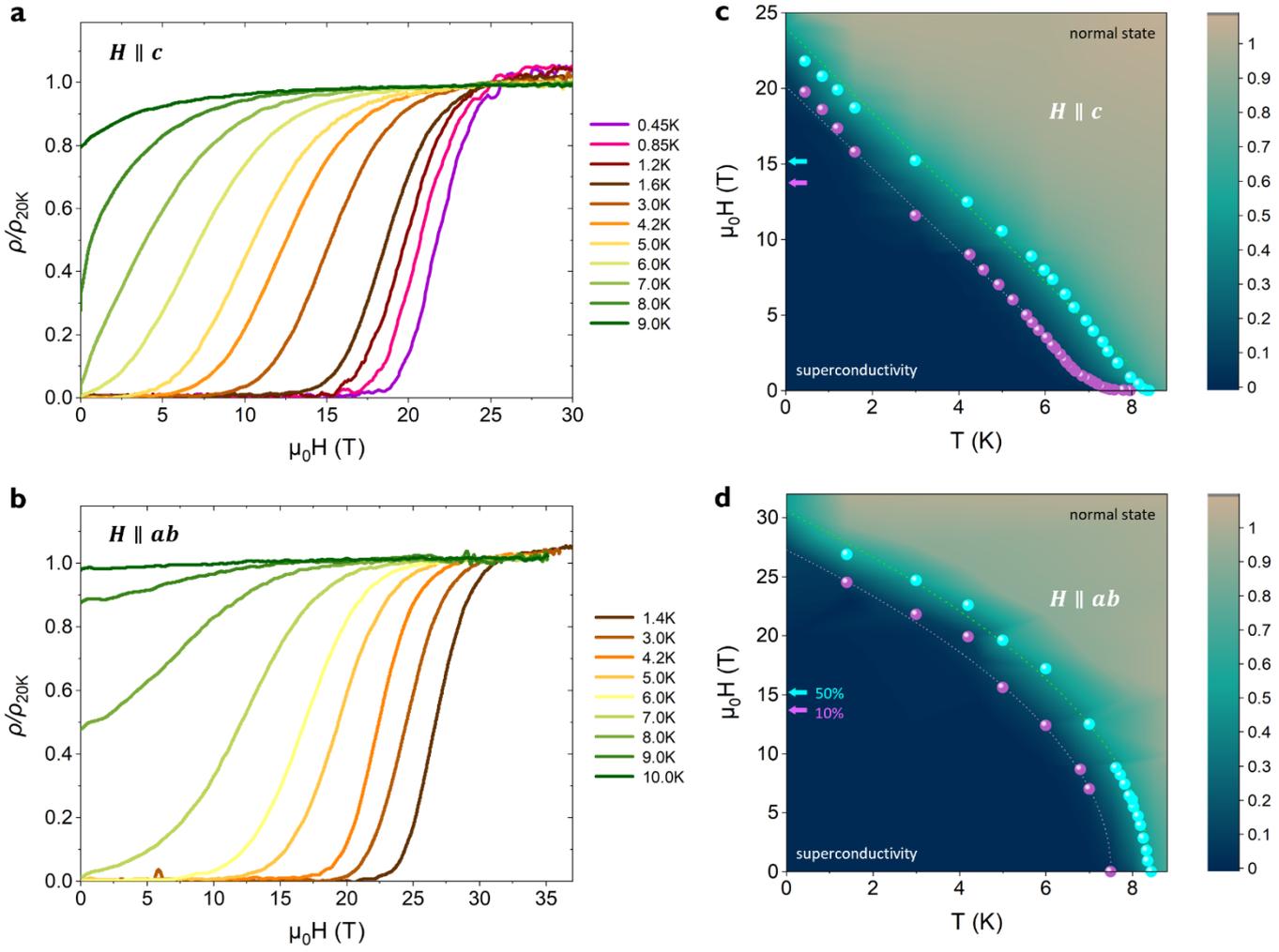

**Figure 2: Large Pauli-limit violation in La$_{0.8}$Ca$_{0.2}$NiO$_2$ thin film. (a-b)** $R - H$ curves measured in pulsed high magnetic fields at various temperatures, where the magnetic field is applied along **(a)** $H \parallel c$ and **(b)** $H \parallel ab$ directions. The transition widths under magnetic fields $\Delta H_{0 \to 90\%}$ are narrower as the temperature decreases down to the lowest temperature. **(c-d)** $H - T$ phase diagrams for **(c)** $H \parallel c$ and **(d)** $H \parallel ab$ directions. The Pauli-limits $H_p = 1.84 \times T_c$ are marked as cyan and purple arrows for 50% resistivity and 10% resistivity thresholds, respectively. The experimental upper critical fields $H_{c2}$ exceeds $H_p$ significantly in both $H \parallel c$ and $H \parallel ab$ directions, even at zero resistivity threshold. The temperature dependence of $H_{c2}(T)$ can be fitted to a linear relation at $H \parallel c$ and $H_{c2}^{ab} \propto \sqrt{1 - \frac{T}{T_c}}$ at $H \parallel ab$ directions, at both 10% and 50% resistivity thresholds, as shown by the dashed lines for eye-guide.

**Pauli-limit violation**

Next, we compare the magnitude of the upper critical fields between different rare-earth ions in the infinite-layer nickelate family. Surprisingly, we noticed that a very large magnetic field is needed to suppress the superconductivity (at zero-resistivity threshold) in the La- system, ~20 T field (for $T_{c,0} \sim 7.3$ K) as compared to a much smaller field (~8 T) in a higher $T_{c,0} \sim 9$ K Nd- sample ($H \parallel c$, see **Fig. S1h**), effectively a $H_{c2}/T_c$ that is three times higher. We measured the upper critical fields of Nd-nickelate thin films down to 0.03 K (see **Fig. 3(left)**) and observed that, at all doping levels $x$, both $H_{c2}^c$ and $H_{c2}^{ab}$ are smaller than the Pauli (Clogston–Chandrasekhar)-limit[1,2] $H_p = 1.84$ T/K $\times T_c$, for a weak-coupled BCS superconductor (**Fig. 3** and **Fig. 4**). This is in agreement to the report on Nd$_{0.775}$Sr$_{0.225}$NiO$_2$ which[11] proposed the first discovered superconducting member of nickelates, that Nd$_{1-x}$Sr$_x$NiO$_2$, is a spin-singlet superconductor with even parity like cuprate. On the other hand, as shown in **Fig. 2** and **Fig. 3**, the upper critical fields of the La$_{0.8}$Ca$_{0.2}$NiO$_2$ are much larger than the Pauli-limit for a spin-singlet superconductor in every direction of the applied magnetic field, even when using a small 10% resistivity threshold. We investigated many samples of different hole doping $x$ with different dopants Ca$^{2+}$ or Sr$^{2+}$ in the La-system. All of them undoubtedly show a violation of the Pauli-limit (**Fig. 3** and **Fig. 4**) in all directions. The measured $H_{c2}^{ab}$ well exceeds $H_p$ at temperature as high as $0.88 \times T_c$, while $H_{c2}^c > H_p$ at $0.59 \times T_c$ in the largest violation of the Pauli-limit observed (**Fig. 3(right)**).

Such a strong violation at temperature close to $T_c$ suggests that the violation of Pauli-limit is not a consequence of an unknown second superconducting phase at low temperature nor related to the Fulde-Ferrell-Larkin-Ovchinnikov (FFLO) state.[12] The violation of the Pauli-limit occurs in the entire doping dependent superconducting phase diagram of the La$_{1-x}$(Ca/Sr)$_x$NiO$_2$ system, and therefore, should not be related to a possible large pseudogap. To our best knowledge, the observed large (2 – 3 times) violation of the Pauli-limit in all crystallographic directions (**Figs. 4b,c**) is unlikely to be accounted for by the strong-coupling mechanism, where the superconducting gap ratio $\frac{\Delta(0)}{k_B T_c}$ is large and up to twice of the weak-coupling values. On top of an unphysically large value of $\Delta_{La}$ is needed to explain the large violation of Pauli-limit in La-nickelate as compared to the Pauli-limited $H_{c2}$ in Nd-nickelate, such an enormous violation is unheralded in a strong-coupling superconductor without additional enhancement by other mechanism such as strong spin-orbit-coupling.[34] For example, in cuprates, YBa$_2$Cu$_3$O$_{7-\delta}$ ($T_c \sim 93$ K) is a strong-coupled $d$-wave superconductor with $H_{c2}^{\perp}(0) \sim 120$ T,[35,36] that is smaller than the Pauli-limit $H_p = (1.84 \times 93)$ T $\approx 171$ T, while $H_{c2}^{\parallel}$ is drastically enhanced to $\sim 1.4$ times $H_p$ by large spin-orbit coupling.[35]

A large violation of the Pauli-limit can happen on a spin-singlet superconductor, typically only when the magnetic field is applied along some of the crystallographic directions, due to a strong spin-orbit-coupling (SOC),[13,14] such as in a two-dimensional system with Ising superconductivity.[14] The strong SOC imposes an additional effective Zeeman field which locks the spins of Cooper pairs at a particular direction, for example, in the out-of-plane direction, which prevents the applied magnetic field along the (orthogonal) in-plane direction from easily flipping the spins and, therefore, resulting in a substantial Pauli-limit violation for $H_{c2}$ along the (orthogonal) in-plane direction. In another example of strong Rashba SOC in the noncentrosymmetric superconductors,[13] the spins are locked in the in-plane direction and consequently, boosting the $H_{c2}$ in out-of-plane direction. However, the upper critical field along at least one of the directions, like perpendicular to the plane in the Ising superconductor will still be smaller than the Pauli-limit for a spin-singlet superconductor.[14] In La-nickelate, we observed around two times (up to about three times for $H_{c2}^{ab}$) violation of Pauli-limit in both $H \parallel ab$ and $H \parallel c$ directions (see **Figs. 4b,c**). Additionally, there is no experimental evidence suggesting a strong SOC in the infinite-layer nickelate that is distinctive between different rare-earth ions.

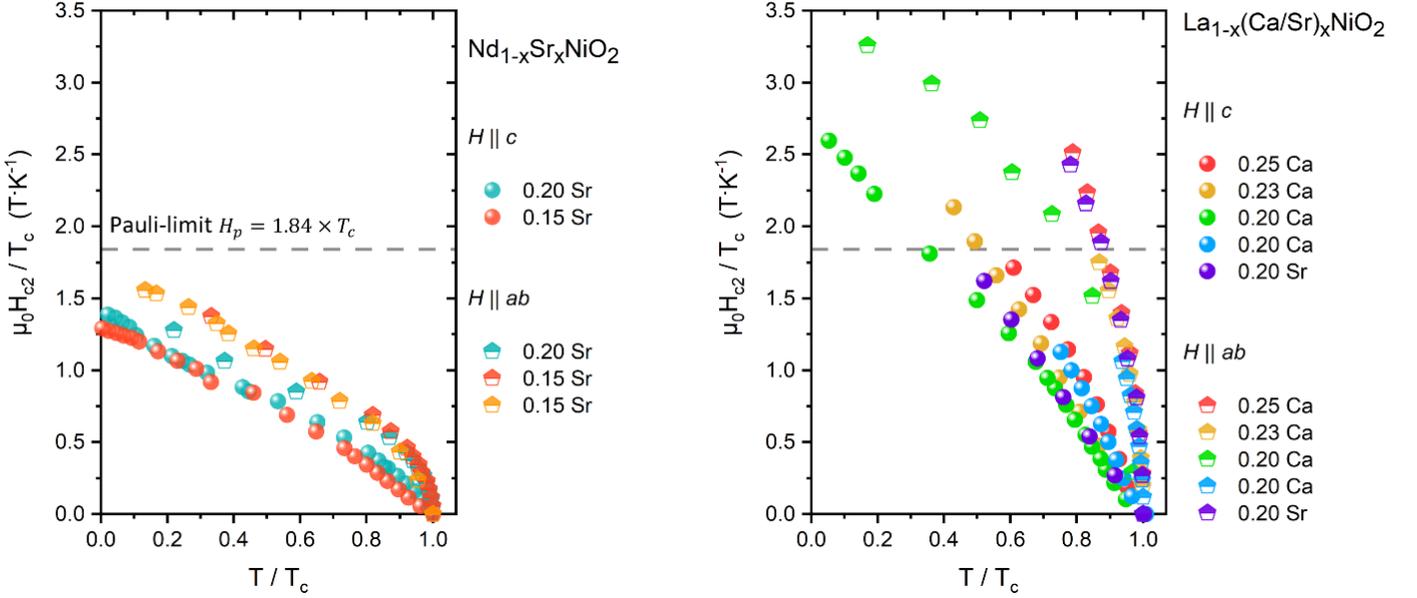

**Figure 3: Pauli-limit violation in the family of La- nickelate.** $H_{c2} - T$ phase diagram of the Nd-nickelate $Nd_{1-x}Sr_xNiO_2$, and La-nickelate $La_{1-x}Ca_xNiO_2$ and $La_{1-x}Sr_xNiO_2$ plotted using 50% resistivity threshold and normalised to $T_c$. The Pauli-limit for a spin-singlet superconductor ($H_{c2} < H_p$) is indicated as the horizontal dashed line at $\frac{\mu_0 H_{c2}}{T_c} = 1.84$. The magnetic fields are applied along the $c$-axis (circle) and $ab$-axis (pentagon) directions. For Nd-nickelate $Nd_{1-x}Sr_xNiO_2$ thin films **(left)**, the upper critical fields in both $H \parallel c$ and $H \parallel ab$ directions are smaller than $H_p$. A larger 0.2 Sr doped film appears to be more isotropic than the lower 0.15 Sr doped film. No strong sign of 'anomalous upturn' at low temperature is seen, even at temperature down to 0.03 K for $Nd_{0.85}Sr_{0.15}NiO_2$. For La-nickelate $La_{1-x}Ca_xNiO_2$ and $La_{1-x}Sr_xNiO_2$ thin films **(right)**, all measured doping levels ($x = 0.2, 0.23, 0.25$ for Ca-doped) show undoubtfully a violation of the Pauli-limit. The upper critical field $H_{c2,50\%}$ exceeds $H_p$ at temperatures as high as 0.88 $T_c$ for $H \parallel ab$, and 0.59 $T_c$ for $H \parallel c$. Consistent phenomena are seen in every Ca-doped and Sr-doped La- nickelate sample, indicating the observed Pauli-limit violation is universal to La- infinite-layer nickelate.

On the other hand, a spin-triplet pairing state has parallel spin-alignment (even symmetry spin state) and, therefore, is unaffected by the upper bound on $H_{c2}$ imposed by the Zeeman effect (Pauli-limit).[15,37–39] For example, the spin-triplet superconductor UTe$_2$ has $H_{c2}$ values easily exceeding $H_p$ in all three directions, with the smallest violation $\frac{H_{c2}^a(0)}{H_p} \sim 2$ along the $a$-axis.[15] Spin-triplet Cooper pairing has recently attracted profound interest due to the potential application of topological superconductivity in quantum computing and dissipationless spintronics.[16,17] However, the intrinsic triplet superconductors are scarce,[15,37–39] many of which have a low $T_c \sim 1$ K. This severely limits experimental accessibility due to the challenging low-temperature and high-pressure conditions.[15,37–39] We plotted the $H_{c2}(0)$ vs $T_c$ for the existing spin-triplet superconductor family – the Uranium based compound as a comparison to our findings of $H_{c2}$ in the infinite-layer nickelates in this study (see **Fig. 4a**). Most of the U-based superconductors have ferromagnetic interaction coupled with superconductivity, which is believed to be relevant for their spin-triplet pairing mechanism, except for UTe2.[15] UTe2 is a paramagnetic heavy-fermion superconductor with the highest $T_c$ recorded for spin-triplet superconductors, of $T_c \sim 1.6$ K that increases up to ~3 K under high pressure.[37] On the other hand, spin-triplet pairing and time-reversal-symmetry breaking order parameter are also proposed in nonmagnetic superconductors.[40–42] The magnetic ground state of infinite-layer nickelate is an open and controversial topic due to the absence of bulk superconducting samples[43] and challenging material synthesis which results in prone formation of extended defects in the thin-film samples. The recent muon spin relaxation study suggested possible short-range correlated magnetism or ordered magnetism in the La- infinite-layer nickelate thin films.[44] It is presently unclear if a spin-triplet or mixed singlet-triplet pairing mechanism can be constructed for the newfound La1-x(Ca/Sr)xNiO2 nickelate system. However, the attractive 'high-$T_c$' with an onset $T_c \sim 10$ K is around an order of magnitude higher than the existing spin-triplet superconductors of $T_c \sim 1$ K, and highly favourable for widening the application of spin-triplet superconductivity. While complementary evidence such as measurements of spin susceptibility across $T_c$ and the detection of Majorana edges states would be necessary to confirm the spin-triplet nature, the challenging material synthesis and absence of a bulk superconducting sample render these experimental investigations difficult to be performed at present. Nevertheless, the large ~2 times violation of the Pauli-limit in all crystallographic directions in La- infinite-layer nickelate begs further look into this exciting new family of superconductors.

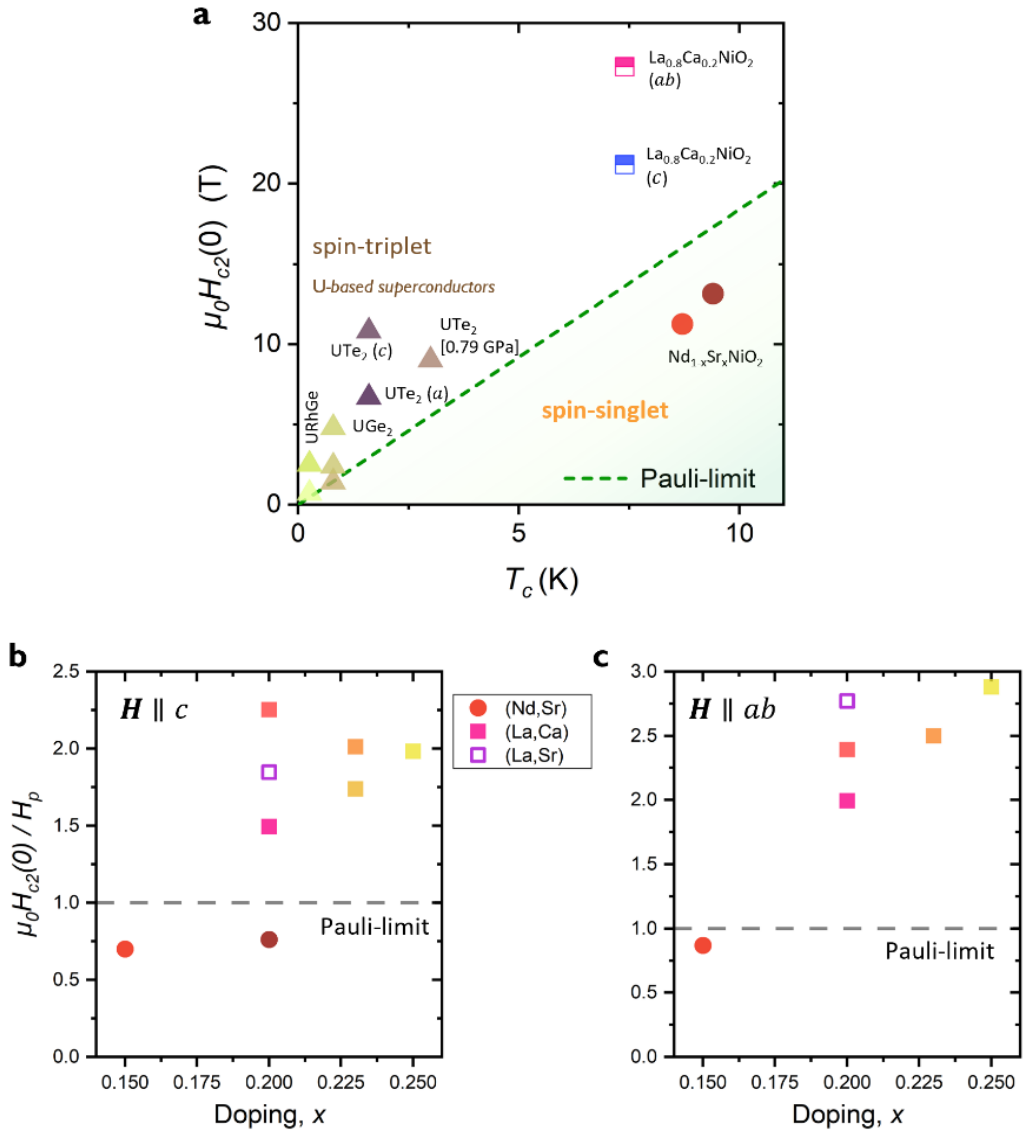

**Figure 4: 'High-temperature' ($T_c \sim 10$ K) violation of Pauli-limit. (a)** Zero-temperature upper critical field $\mu_0 H_{c2}(T = 0 \text{ K})$ as a function of superconducting transition temperature $T_c$ for La- and Nd- nickelates compared to the existing spin-triplet superconductors, $T_c \sim 1.6$ K for UTe$_2$ ($T_c \sim 3$ K at high pressure). U-superconductors data was adapted from Ref.[15,37–39]. The magnetic field was applied along crystallographic axis a-, b- or c-axis. The Pauli-limit for a spin-singlet superconductor, which $H_{c2} < H_p = 1.84 \times T_c$ is shown as the green dashed line. **(b-c)** Estimated Pauli-limit violation ratio $\frac{\mu_0 H_{c2}(T=0 \text{ K})}{H_p}$ as a function of hole doping $x$. For La- nickelates, the $\mu_0 H_{c2}(0)$ is extrapolated **(b)** using a linear fitting for $H \parallel c$ and **(c)** using $H_{c2}^{ab} \propto \sqrt{1 - T/T_c}$ for $H \parallel ab$ [see **Figs. 2c,d**]. Below $\frac{\mu_0 H_{c2}(0)}{H_p} = 1$ line, the upper critical field is smaller than the Pauli-limit, which is expected for a spin-singlet superconductor. For La- nickelates, all samples of various doping $x$ and dopant (Ca/Sr) show a large, around two times violation of Pauli-limit for a spin-singlet superconductor in all directions, suggesting non-spin-singlet Cooper pairing.

# Methods

## Sample growth and preparation

The infinite-layer nickelate $Nd_{1-x}Sr_xNiO_2$ and $La_{1-x}(Ca/Sr)_xNiO_2$ thin films were grown on $SrTiO_3$ (001) substrates using pulsed laser deposition and $CaH_2$ topotactic reduction process under conditions as previously reported.[31]

## Transport measurement under static magnetic fields up to 14 T

The wire connection for the electrical transport measurement was made by Al ultrasonic wire bonding. The transport measurements down to 2 K were performed using a Quantum Design Physical Property Measurement System. We perform additional magnetotransport measurements in a dilution fridge (base temperature ~ 10 mK) equipped with a 14 T static field magnet and a rotator to scan from $H \parallel c$ to $H \parallel ab$ directions.

## High magnetic pulsed-field transport measurement

The magnetotransport measurements under high magnetic fields (55 T pulsed with a duration of 300 ms) were performed in a $^3$He cryostat for measurements down to 0.4 K, and $^4$He cryostat for measurements down to 1.4 K at LNCMI-Toulouse. We apply a DC current excitation of 10 µA. The sample was warmed up to 20 K and then cooled down to the desired temperature before applying a pulsed field shot. For measurements above 4.2 K, the sample is not immersed in the liquid phase of the Helium bath and hence heating from eddy currents during pulsed magnetic field shots can significantly affect the actual temperature of the sample. We used the magnetoresistance measured when the magnetic field increases to minimise temperature inaccuracy from overheating.

**Pauli-limit violation from magnetotransport**

We define the normal state resistivity as the resistivity at 20 K, $\rho_{xx,20K}$ under zero-field. The resistivity values were normalised to $\rho_{xx,20K}$, noted as $\rho/\rho_{20K}$, to indicate the fraction of superconducting transition. The transition points were defined using several resistivity thresholds, which zero-field $T_{c,\,x\%}$ is defined at the temperature at which the resistivity value reaches $x\%$ of normal state resistivity. Similarly, the upper critical fields $\mu_0 H_{c2,\,x\%}(T)$ is defined based on resistivity thresholds. For a selected resistivity threshold, the ratio $\frac{\mu_0 H_{c2}(T)}{T_c}$ is calculated to determine if violation of Pauli-limit for a conventional spin-singlet weak-coupled superconductor occurs, when $\frac{\mu_0 H_{c2}(T)}{T_c} > 1.84$.

**Data availability**

The data that support the findings of this study are available upon reasonable request from the corresponding author.

**Acknowledgement**

We acknowledge discussions with Alexander A. Golubov. This research is supported by the Ministry of Education (MOE), Singapore, under its Tier-2 Academic Research Fund (AcRF), Grant No. MOE-T2EP50121-0018, and Research Grants Council of the Hong Kong SAR, Grant No. A-CUHK 402/19. We acknowledge the support of LNCMI-CNRS, a member of the European Magnetic Field Laboratory (EMFL) under the proposal numbers TMS10-219 and TMS10-221 and the funding support from the European Union's Horizon 2020 research and innovation programme under grant agreement No 823717 - ESTEEM3.

## Authors contribution

AA conceived and led the project. AA, LEC, SWZ, WE, SKG designed the experiments. LEC, SWZ, ZTZ synthesized the infinite-layer nickelate thin films supported by ZSL and ZYL. LEC conducted the electrical measurements down to 2 K with 9 T static fields supported by SWZ, ZTZ, ZSL, SKS. WE, MP, MG performed the high-field magnetotransport experiment in a $^3$He-based refrigerator operating down to 350 mK under pulsed magnetic field of up to 60 T developed by MN and AZ. KYY and SKG performed magnetotransport measurements in a dilution fridge (base temperature ~ 10 mK) with 14 T field. T.H., J.D., P.N., P.A.V.A. conducted the electron microscopy. LEC, SWZ, KYY, SKG, WE, AA analysed the data with input from all authors. LEC and AA wrote the manuscript with advice from all authors.

## Competing Interests

The authors declare no competing interests.

# Extended Data

for

# Pauli-limit violation in lanthanide infinite-layer nickelate superconductors

L. E. Chow,[1] K. Y. Yip,[2] M. Pierre,[3] S. W. Zeng,[1] Z. T. Zhang,[1] T. Heil,[4] J. Deuschle,[4] P. Nandi,[1] S. K. Sudheesh,[1] Z. S. Lim,[1] Z. Y. Luo,[1] M. Nardone,[3] A. Zitouni,[3] P. A. van Aken,[4] M. Goiran,[3] S. K. Goh,[2] W. Escoffier,[3] A. Ariando[1,✉]

[1]*Department of Physics, Faculty of Science, National University of Singapore, Singapore 117551, Singapore*

[2]*Department of Physics, The Chinese University of Hong Kong, Shatin N.T., Hong Kong SAR, China*

[3]*LNCMI, Université de Toulouse, CNRS, INSA, UPS, EMFL, 31400 Toulouse, France*

[4]*Max Planck Institute for Solid State Research, 70569 Stuttgart, Germany*

✉To whom correspondence should be addressed: ariando@nus.edu.sg

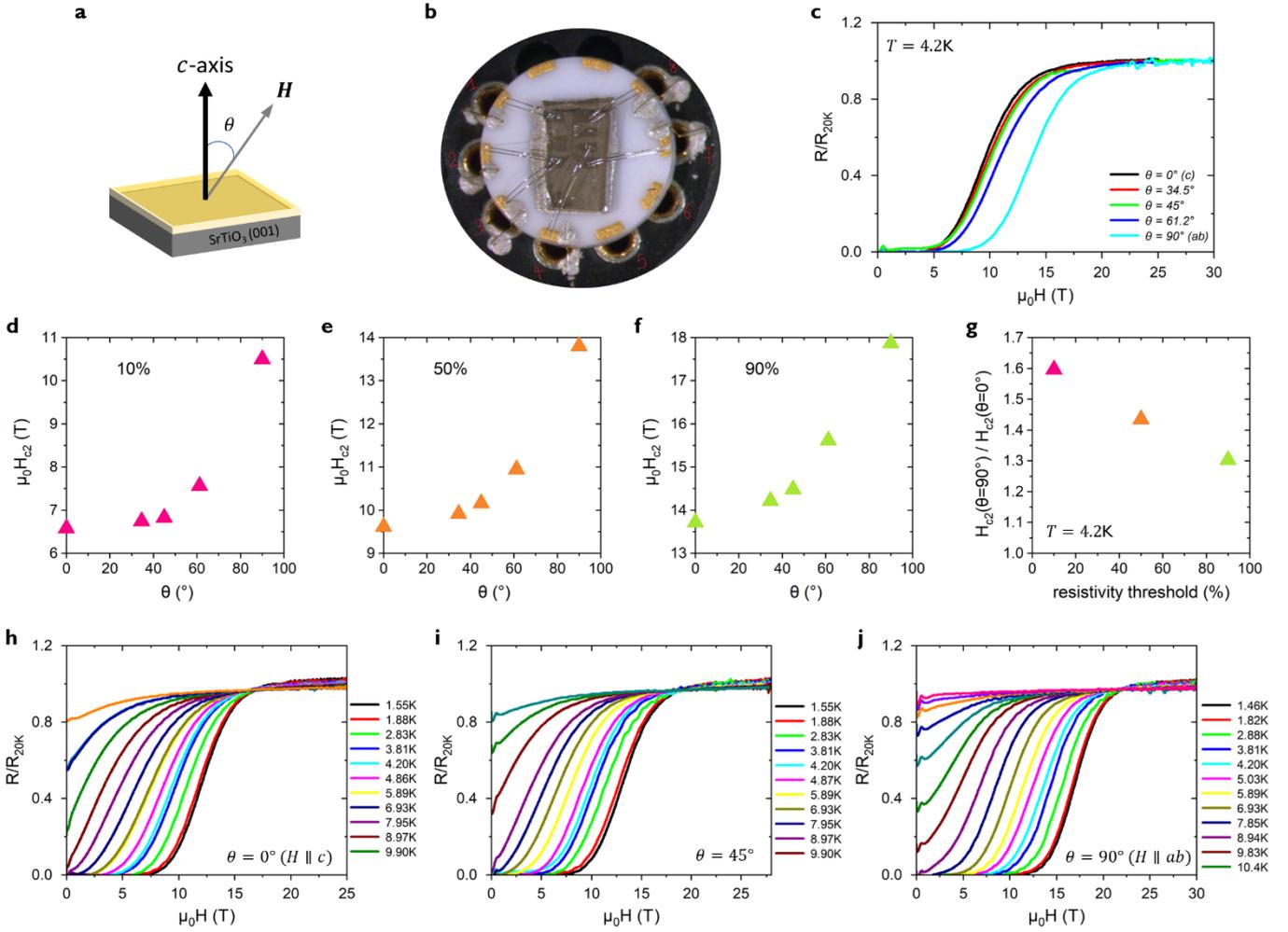

**Figure S1: Magnetotransport measurements of the upper critical fields of Nd$_{0.85}$Sr$_{0.15}$NiO$_2$ thin film using pulsed magnetic fields.** **(a)** A schematic view of the orientation of the magnetic field direction. **(b)** Photo of a sample to be inserted into cryostat for high magnetic pulsed fields measurements. **(c)** $R - H$ curves at 4.2 K for magnetic fields applied along several $\theta$ directions. **(d-f)** Upper critical field $H_{c2}$ as a function of polar angle $\theta$ at 4.2 K measured using several resistivity thresholds **(d)** 10%, **(e)** 50%, **(f)** 90%. **(g)** Ratio of $\frac{H_{c2}(\theta=90°)}{H_{c2}(\theta=0°)}$ as a function of resistivity threshold % at 4.2 K, where 100% normal state resistivity is defined as zero-field resistivity at 20K. **(h-j)** $R - H$ curves measured at several temperatures below 20K for magnetic fields applied at **(h)** $\theta = 0°$, **(i)** $\theta = 45°$, **(j)** $\theta = 90°$. (The 'tiny bump' in resistivity near $H = 0$ T is affected by the parasitic induced voltage but not a real feature.)

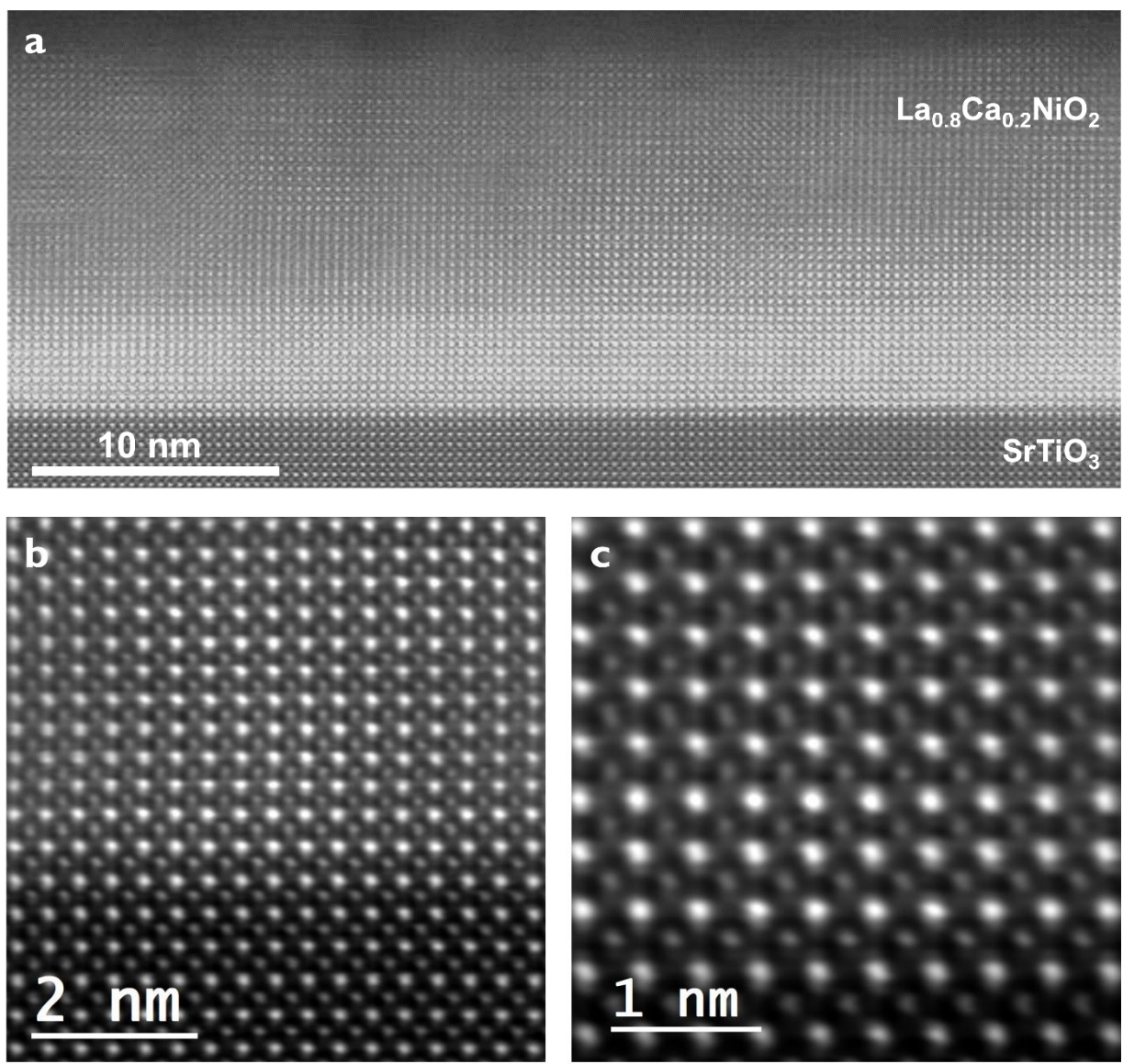

**Figure S2: Cross-sectional scanning transmission electron microscopy high-angle annular dark field (STEM-HAADF) image of the superconducting La$_{0.8}$Ca$_{0.2}$NiO$_2$ thin film of ~15 nm thick. (a)** A large area of the infinite-layer thin film. **(b-c)** High-magnification images of the film-substrate interface. A radial Wiener filter was used to reduce the noise of the HAADF images.